\documentclass[aps,twocolumn,showpacs,superscriptaddress,groupedaddress, nofootinbib]{revtex4-1}

\usepackage{hyperref}
\usepackage{multirow}
\usepackage{mathrsfs,amsmath} 
\usepackage{verbatim}
\usepackage{amsmath}
\usepackage{amsfonts}
\usepackage{amssymb}
\usepackage{amsthm}
\usepackage{bm}
\usepackage{xparse}
\usepackage{xcolor}
\usepackage{textcase}
\usepackage{url}
\usepackage{lipsum}
\usepackage{appendix}
\usepackage{dsfont}
\usepackage{epstopdf}
\usepackage{footnote}
\usepackage{tgbonum}
\usepackage{blindtext}
\usepackage{enumitem}
\usepackage{mathrsfs,amsmath} 
\usepackage{float}
\usepackage{subfloat}
\usepackage{wrapfig}
\usepackage{pslatex}
\usepackage{amsmath}
\usepackage{amsfonts}
\usepackage{setspace}
\usepackage{epstopdf}
\usepackage{wrapfig}
\usepackage{csquotes}
\usepackage{hyperref}
\usepackage{graphicx}
\usepackage{amssymb}
\usepackage{multirow}
\usepackage{xcolor}
\usepackage{framed}
\usepackage{mathtools}
\usepackage{tcolorbox}
\definecolor{shadecolor}{rgb}{0.8,0.9,1}
\DeclareDocumentCommand{\Tr}{m m O{\big}}{{\rm Tr}_{\:\!{#1}}#3({#2}#3)}

\begin{document}
\title{The open past in an indeterministic physics}
\author{Flavio Del Santo}
\affiliation{
Institute for Quantum Optics and Quantum Information (IQOQI-Vienna), A-1090 Vienna, Austria; and
Faculty of Physics, University of Vienna, A-1090 Vienna, Austria}
\author{Nicolas Gisin}
\affiliation{Group of Applied Physics, University of Geneva, 1211 Geneva 4, Switzerland; and Schaffhausen Institute of Technology (SIT), Geneva, Switzerland}

\date{\today}

\begin{abstract}
Discussions on indeterminism in physics focus on the possibility of an open future, i.e. the possibility of having potential alternative future events, the realisation of one of which is not fully determined by the present state of affairs. Yet, can indeterminism affect also the past, making it open as well? We show that by upholding principles of finiteness of information one can entail such a possibility. We provide a toy model that shows how the past could be fundamentally indeterminate, while also explaining the intuitive (and observed) asymmetry between the past --which can be remembered, at least partially-- and the future --which is impossible to fully predict.

\end{abstract}

\maketitle
%%%%%%%%%%%%%%%%%%%%%%%%%%

%%%%%%%%%%%%%%%%%%%%%%%%%%%%%%%%%%%%%%%%%
\section{Introduction}
One of the fundamental questions in the philosophy of physics aims to establish whether the evolution of every physical system is governed by determinism --i.e. if any two states of a system at different times are related by a one-to-one connection-- or if future events can in principle have multiple potential outcomes that are not fully determined by the past states. While for centuries it was believed that such a deterministic account is uncontroversially true (mostly due to the empirical success of the equations of motion of classical physics and their mathematical properties),\footnote{Note that there are exceptional cases where the classical equations of motions do not have a unique (i.e. deterministic) solution, such as in the case of Norton's dome \cite{norton}.} the advent of quantum physics raised a lasting debate that leans towards the solution that the world is actually indeterministic. In recent years, we have proposed that even classical physics should be interpreted indeterministically, if one assumes the reasonable principle of finiteness of information density (i.e., finite volumes of space can only contain a finite amount of information) \cite{gisin1, delsantogisin1, NGHiddenReals, delsanto2021, delsantogisin2}. 

Indeterminism certainly grants to the direction of time a fundamental role, for changes really happen: while time passes, from an array of mutually exclusive potential events, only one obtains. However, once one event has been actualized, it is considered determinate and this status does not undergo further change as time passes. While the future is open in an indeterministic world, the past is there to stay, as encapsulated also by C. F. von Weizs\"acker when stating that ``the past is factual, the future is possible'' \cite{weizsacker}. Nevertheless, some authors in the philosophical literature (notably, J. \L ukasiewicz \cite{luka} and M. Dummett \cite{dummett68, dummett04})  have hinted at the possibility that upholding indeterminism could have consequences for the past too. Yet, this idea that the past could also not be fully determined by the present state of affairs at a fundamental level in an indeterministic world seems to have been often overlooked even within the numerous debates about indeterminism. 

It is the aim of this paper to revive the question: \emph{can the past be open?}, hoping that this will motivate further attempts of addressing this problem. It should immediately be noticed that while we are in general not able to predict the future with certainty --which leaves room for a philosophical debate on whether to interpret this as either a lack of knowledge about the underlying determinism or as fundamental indeterminism-- we do seem to remember the past (that is, \emph{it exists} information about the past that can in principle be remembered with certainty).  This, therefore, begs the consequent question: \emph{if the past is open, why do we observe such an asymmetry in terms of predictions retrodictions?} In what follows, building further on principles of finiteness of information that entail indeterminism, we provide a toy model that aims to show how the past as well could become (again) fundamentally indeterminate while explaining the intuitive asymmetry between the past, which seems determinate in our recollections, and the future, which is in general impossible to fully predict.

\section{Can the past be open?}
In the most general sense, without any ontological commitment, the \emph{laws of physics} can be regarded as causal relations that connect \emph{pure states} of a system at different instants of time (given a suitable conception of a time-arrow). Note that a pure state is in this case generally defined as the state encapsulating the maximal amount of information existing at present about each relevant degree of freedom of the system itself. Note that, as in our previous works \cite{gisin1, delsantogisin1, NGHiddenReals, delsanto2021, delsantogisin2},  we uphold a physical interpretation of information. This means that information becomes meaningful only when embodied in distinguishable states of a physical system. In the words of Landauer who pioneered this ideas, ``Information is not a disembodied abstract entity; it
is always tied to a physical representation. It is represented by engraving on a stone tablet, a spin, a charge,
a hole in a punched card, a mark on paper, or some
other equivalent. This ties the handling of information to all the possibilities and restrictions of our real
physical word, its laws of physics and its storehouse
of available parts.'' \cite{landauer}.\footnote{This view departs from other popular views in the philosophy of information that regard information as an epistemic or even an abstract concept (see \cite{lombardi, floridi}).}

 In the broadest possible sense, two events --named the cause ($C$) and the effect ($E$), respectively-- are said to be causally related in the case that if $C$ obtains, it influences the tendency for $E$ to obtain. Note that for this to happen, $C$ must happen before $E$. Following this general definition of causal laws, in the present note we consider different examples of indeterministic physics, showing that indeterminism can entail not only the known indeterminacy about the future but about the past as well, which truly evaporates as time passes. By this we do not mean merely an epistemic statement, namely that it is \emph{our knowledge} of the past which evaporates. Rather, it is the \emph{available information} that disappears from some degrees of freedom of the universe that were encoding it, and therefore it ceases to exist altogether.
 
\subsection{Deterministic causality and spontaneous ``acausal'' events}
In ``orthodox'' classical physics, it is assumed that causally connected events are all related by necessity (deterministic causality). The causal connections are in this way fixed for all times,  i.e., every event that obtains is the certain result of the previous events that obtained before, without the possibility of alternative ones. This forms a causal ``chain'' all given at once, \emph{ad infinitum} towards both the past and the future.

However, possible alternatives have been already considered in the vast literature devoted to (in)determinism. As a first example, consider the following possibility to avoid a deterministic world: even if one maintains deterministic causality as the norm, some authors had it that it is possible to conceive that the chain of necessary cause-effect relations is from time to time interrupted by an ``acausal'' event, that spontaneously obtains independently of any other previous event. Note that this is not the position that we endorse, for the concept of a genuinely ``acausal'' event that suddenly arises out of nothing seems arbitrary and unsatisfactory. More formally, if the events that are one the cause of the other are represented as nodes in a directed (because causes precede their effects), acyclic (because an event is never the cause of itself) graph, spontaneous acausal events may create disconnected parts in the graph. This possibility clearly leads to the openness of the future in so far as some future events may not be inferred even given complete determinacy of all the events that are causally connected to the present (and to the past), because a possible future spontaneous event can create a new unpredictable, disconnected subgraph of the whole causal graph. In a similar fashion, \L ukasiewicz suggested that the same conclusion should be reached about the past: if no causal connections still exist at present with a certain past event (or more realistically a whole past, disconnected subgraph), that event should not \emph{any more} be considered determined. In his words, ``facts whose effects have disappeared altogether, and which even an omniscient mind could not infer from those now occurring, belong to the realm of possibility and not the realm of actuality. One cannot say about them that they took place, but only that they were \emph{possible}.'' \cite{luka}.
In what follows, we would not further consider this somewhat unsatisfactory approach of ``disconnected'' deterministic chains, but rather develop a framework of causally complete and yet fundamentally indeterministic physics.  

\subsection{Probabilistic causality and finite information}
Since the mid-twentieth century, philosophers of science such as of K.R.  Popper \cite{popper}, J. Earman \cite{earman}, W. Salmon \cite{salmon}, P. Dowe \cite{dowe}, H. Reichenbach \cite{reichenbach}, I. J. Good \cite{good} and P. Suppes \cite{suppes2} have generalized the concept of causality to indeterministic scenarios, developing the idea of \emph{probabilistic causality}. This states that an event $C$ affects the tendency of another event $E$  to happen, but E is not bound to happen by necessity. This generalizes the Leibnitzian principle of sufficient reason: “there is nothing without a reason, or no effect without a cause”, but causation is here meant as a looser connection between events than determinism. The standard way to quantify a tendency for an event to happen is to make use of probabilities. In this case, however, probabilities are thought of as propensities (as introduced by Popper \cite{popper}), i.e. objective properties of a physical system, in relation to an idealized environment, to change from a pure state to another (see also \cite{gisinprop1, gisinprop2}).\footnote{Note that in our view, which slightly departs from Popper's original one, we refer to an idealized environment, because a real environment would come with its own noise that may contribute to the frequencies of outcomes but this is irrelevant to the system's propensities. This can be understood by looking at quantum theory where the propensities are fully characterized by the pure state and the idealized measurement operators.}

\begin{figure}[!]
\begin{center}
\includegraphics[width=6cm]{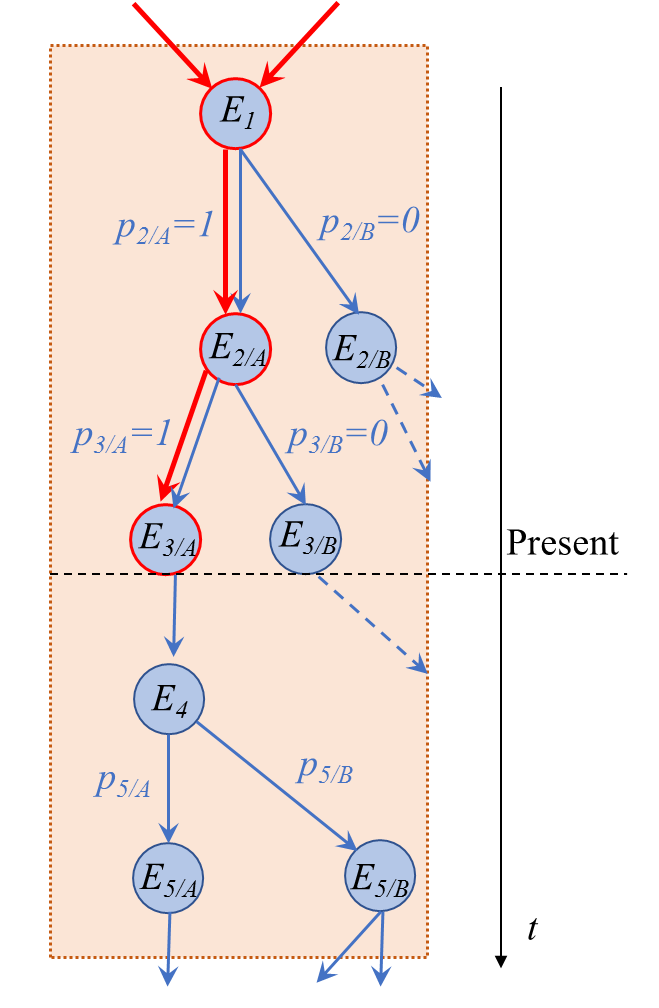}
\caption{\small \textbf{Multigraph representing causal indeterministic connections}. The nodes represent events, the red edges connect events that have already obtained, while the blue edges represent potential causal connections, associated with a weight that represents the propensity for an event to obtain. The dotted arrows represent causal connections about potential futures that will never be because the propensity of realizing that branch has already evolved to zero.}
\label{graph}
\end{center}
\end{figure}

In a probabilistic causal scenario, events can be graphically represented as embedded in a directed and acyclic multigraph, with two types of edges (Fig. \ref{graph}). The first type of edges represents a ``potential'' causal connection and each of these edges is weighted with the propensity that relates two events, i.e. associated with a non-negative rational number such that the sum of the weights connecting a cause to all possible, mutually exclusive events is equal to 1. The second type of edges represents instead an ``actual'' causal connection, picking only one among the mutually exclusive ``potential'' causal connections (these edges can also be regarded as each weighted with a constant propensity equal to 1, which corresponds to certainty).
The potential graph represents all the possible alternative histories and it seems \emph{prima facie} independent of the time that one is considering (always given that the information is finite). However, the values of the weights (i.e. the propensities) evolve in time.\footnote{It should be stressed that, contrarily to the deterministic causality discussed above, in a potential causal graph there are no disconnected subgraphs, so the argument of \L ukasiewicz does not seem to be applicable.}  The graph of actual causal connections, instead, develops in time out of the former and it is only when the propensities evolve in time taking the value 1 (thereby ruling out the potential alternative effects to which the corresponding propensities all take the value 0) that this can be constructed. Note that this implicitly defines the concept of present, which is the time slice that separates the subgraph containing exclusively events connected by propensities 0 or 1 (i.e. the past) from the subgraph that contains arbitrary propensities (i.e. the future).

A clarification on the concept of ``event'' seems in order. In standard probability theory, an event is an element of a sample space, which together with a $\sigma$-algebra and a measure function defines a probability space. Here, we do not necessarily assume that propensities are probabilities at the formal level (i.e. that they fulfill all the Kolmogorov axioms), although they intuitively play the same role in quantifying likelihood. Hence, when we refer to an event, we mean anything that the theory can talk about, namely any proposition to which it is possible to attach a measure of likelihood.\footnote{For example, the proposition “it is raining” can be true without all involved air-particles having fully determined positions and momenta. This indeterminacy implies that a proposition about future weather is presently undetermined. When time passes, particles gain some determinacy (but not full determinacy) and thus some propositions about the future become either true or false.} In particular, one can relate the current theoretical framework in terms of events to our proposed indeterministic interpretation of classical physics based on finite information quantities (FIQs) \cite{delsantogisin1}. A FIQ is an array of propensities that quantify the tendency of each digit of a physical variable (written in binary base) to take the value 1, under the constraint that the total information content of the array is finite. An event can in this context be thought of as the actualization of a digit of a FIQ. 

 %%%%%%%%%%%%%%%%%%%%%%%%%%
%%%%%%%%%%%%%%%%%%%%%%%%%%

It should be clear at this point that a probabilistic view on causality entails indeterminism in  the sense that empirical statements about the future are in general undecidable (they are neither true nor false), not even in principle, despite having complete knowledge of \emph{all} the existing information of the present (pure) states. This makes the law of the excluded middle fail, relating the problem of indeterminism and undecidability of empirical statements to mathematical intuitionism (see \cite{intuition1, intuition2, delsantogisin2}). In this way, propositions the likes of ``it will rain in Paris in a year from now'', or ``the position of a certain particle will be on the left side of a plane at a future time $t_f$'' do not have (at present) a determinate truth value, not only because this might be unknown, but because it is not (yet) determined.

M. Dummett pointed out that this kind of fundamental indeterminacy (which he misleadingly called anti-realism at first \cite{dummett68})  is not only a feature of future events, but it extends towards the past as well when no trace is left that would allow one to decide the truth value of a proposition about the past: ``only those statements about the past are true whose assertion would be justified in the light of what is now the case. [T]his means that there is no one past history of the world: every possible history compatible with what is now the case stands on an equal footing.'' \cite{dummett68} (see also, \cite{dummett04}).\footnote{While supporting the view that statements are true only if there is an empirical justification or verification (a position that he named firstly \emph{anti-realism} and later \emph{justificationsm} \cite{dummett05}), Dummett considered the openness of the past an undesirable consequence.} On this note, we would like to point out that Dummet's misleading term ``anti-realism'' was supposed to actually refer to indeterminism (about the future and in this case about the past too). This confusion is quite widespread, as one of us pointed out: ``Sometimes realism is defined as the hypothesis that every physical quantity always has a value. But then, either this value is unaccessible, hence unphysical, or this value can be revealed by appropriate measurements (to arbitrary good approximation, at least in principle). Hence, these measurements have predetermined outcomes and realism is nothing but a fancy word for determinism'' \cite{gisinreal1}. In fact, our proposal in terms of objective propensities maintains realism while introducing irreducible indeterminism (see also \cite{gisinreal2} for a  more precise definition of indeterministic realism).

Similar considerations to the ones hinted at by Dummet can be refined by making use of the concept of information and its relation to physics. Within probabilistic causality, in fact, the state of a system can thus be regarded as the piece of information recording the whole multigraph $G$ (see a simple illustration thereof in Fig. \ref{graph}), i.e. the ordered collection
%%%%%%%%%%%%%%%%%%%%%%%%%%
\begin{equation}
G(t): [E(t), \Omega(t), P(t)],
\end{equation}
%%%%%%%%%%%%%%%%%%%%%%%%%%
 where $E(t)$ is the set of all events (nodes), $\Omega(t)$ is the set of all potential causal connections (edges), respectively, stored in some degrees of freedom in the universe, while $P(t)$ is the set of propensities (weights) all at the considered (present) time $t$. Note that $G(t)$ encompasses both kinds of edges, those corresponding to the ``potential'' causal influences of the future (weighted blue edges in Fig. \ref{graph}), whose propensities are a rational number between 0 and 1, and all the ``actual'' causal influences that pertain to the past, whose propensities are all either 0 or 1 (red edges in Fig. \ref{graph}).

In a series of papers \cite{gisin1, delsantogisin1, NGHiddenReals, delsanto2021, delsantogisin2}, we introduced the \emph{principle of finiteness of information density}, according to which finite regions of space
can only contain finite information, and showed that this entails the possibility of fundamental indeterminism both in classical mechanics and in special relativity. Along the same lines, we also impose in the present framework that the state  $G(t)$ must contain only finite information. The information content of the state $G(t)$ can be evaluated by, for instance, the Kolmogorov complexity of the string constructed by writing all the elements of the sets that compose $G(t)$, or any other reasonable measure of information. Let us call the information content of the state $I(G(t))= I(E(t))+I(\Omega(t))+I(P(t))$. Imposing that this has a finite value, in general rules out the possibility that the graph has an infinite amount of nodes and edges in any finite time interval, and that any of its elements is represented by a real number (this is why we defined propensities to be rational numbers, but they could more generally be computable numbers which still contain a finite amount of information). Moreover, one can separate the information content of the graph $G(t)$, into a part relative to the past and one relative to the future:
%%%%%%%%%%%%%%%%%%%%%%%%%%
\begin{equation}
I(G(t))= I(G(t))_{past}+I(G(t))_{fut}, 
\end{equation}
%%%%%%%%%%%%%%%%%%%%%%%%%%
where $I(G(t))_{past}= I(E(t))_{past}+I(\Omega(t))_{past}+I(P(t))_{past}$ and $I(G(t))_{fut} =  I(E(t))_{fut}+I(\Omega(t))_{fut}+I(P(t))_{fut}$. While there is no reason that the amount of events as well as of potential causal connections in the past is higher than that in the future or vice versa, there is a clear asymmetry in the information contents $I(P(t))_{past}$ and $I(P(t))_{fut}$. Indeed, while $P(t)_{past}$ is a string containing only 0s and 1s, $P(t)_{fut}$ is a string that can contain any arbitrarily long rational numbers. Therefore, $I(P(t))_{past}	< I(P(t))_{fut}$, which entails 
%%%%%%%%%%%%%%%%%%%%%%%%%%
\begin{equation}
I(G(t))_{past} <I(G(t))_{fut},
\end{equation}
%%%%%%%%%%%%%%%%%%%%%%%%%%
where we have assumed that the amount of events and potential causal connections in the past and in the future is roughly the same. So, describing the future requires storing more information, because this contains all the non-trivial propensities associated with the ``potential'' alternative events (most of which will never obtain). On the other hand, a record of the past history, i.e. of all and only the events that have happened with certainty (i.e. with an updated propensity of 1), clearly requires less information.

Let us now assume that:
\begin{enumerate}[label=(\roman*)]
\item The universe is finite, i.e., it has finite resources occupying a finite volume of space. This, together with the principle of finiteness of information density recalled above, means that the total amount of information in the universe is also finite. Suppose that this total amount of information storable in the universe is upper bounded by $N$ bits. 

\item By homogeneity of time, this maximal amount of storable information in the universe is roughly symmetrically centered around the present instant (as defined above). Namely, about $N/2$ bits are used to record information about the past, i.e. $I(G(t))_{past}$, and roughly the other $N/2$ are recording $I(G(t))_{fut}$. 
\end{enumerate}

 %%%%%%%%%%%%%%%%%%%%%%%%%%
\begin{figure}[]
\begin{center}
\includegraphics[width=6cm]{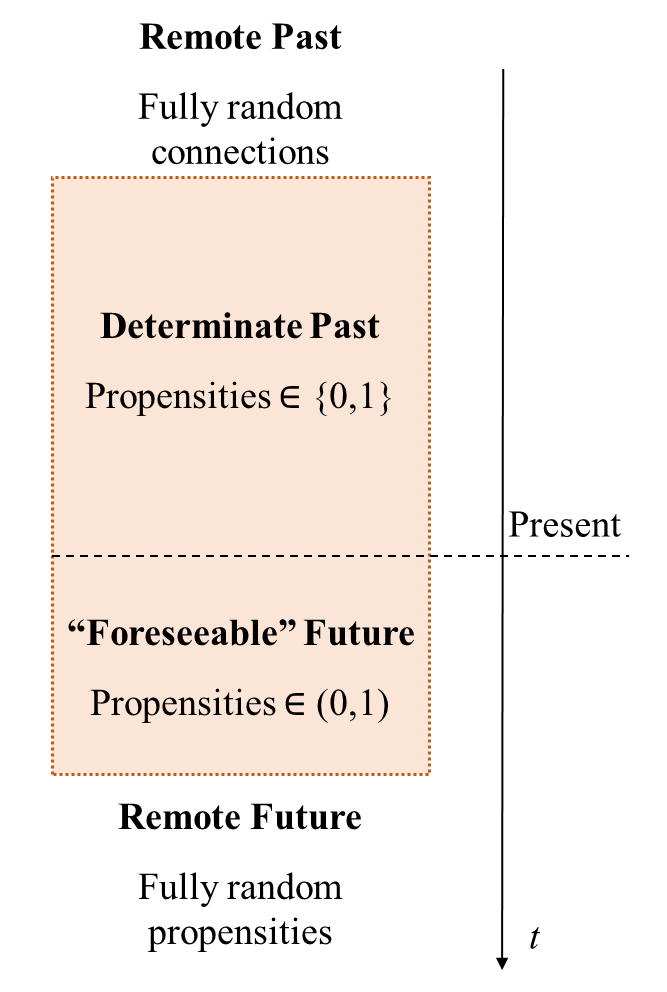}
\caption{\small \textbf{Schematic representation of the different time slices}. Both the remote past and the remote future are characterized by totally random propensities, i.e. each of $n$ mutually exclusive events has an associated propensity of $1/n$ (for the past, we call these ``random connections'' rather than propensities because they do not undergo further evolution). The colored box represents the (non-trivial) state of the universe: the determined past (characterized by actualized events, i.e. with propensities of either 0 or 1) and the foreseeable future (where the propensities are between 0 and 1, but in general not totally random).}
\label{pastfut}
\end{center}
\end{figure}
%%%%%%%%%%%%%%%%%%%%%%%%%%
%even if the events can be arbitrarily close to each other in time, there is only a certain time interval around the present instant about which one can record this information.
Since the amount of information is bounded, while time passes the finiteness of the information requires some past information to be erased. This would mean that the past as well could become open again as the amount of information of which event caused the earliest in time recorded event forces it to be erased. Having lost information about the propensities of the past event, it remains only to assign equal weight to the alternative ``potential'' past events (which we may call ``random past causal connections'' rather than propensities because they do not evolve anymore and never actualize, see Fig. \ref{pastfut}). This leads to the concept of open past, that is, not all the propositions formulated at present about past times are determinate; for instance, the question ``\emph{was} the event E at time $t_j$ caused by the event C at time $t_i < t_j$?'' is in general neither true nor false for times $t_i$ sufficiently far away in the past (i.e. for some $t_i<< t_0$, where $t_0$ is the present time). Note, however, that this construction maintains the desirable asymmetry between past and future in a fundamentally indeterministic theory.\footnote{This thus departs from the proposal to Ref. \cite{markosian}, in which it is argued that the openness of the past results from both indeterministic and time-symmetric nature of the laws of physics.} Both the past and the future are thus indeterminate, and yet we can know more about the past and less about the future as our intuition and experience suggest. Note that this is in agreement with what also Dummet had in mind (and, although with another basic construction, \L ukasiewicz too), namely that the set of past events remain fixed and once time passes the propensities associated with a collection of $n$ mutually exclusive events become homogeneous, i.e. $1/n$.\footnote{The transition from a determinate past to a fully random connection can happen as a sudden jump or have a smother transition that passes through intermediate biased random connections. We leave this distinction as an open question in our model.} Since which one of the possible events has happened is \emph{now} totally random, the information content is zero and this remote past can be considered fundamentally erased, in the same sense that the remote future (i.e. those events that have a totally random propensity associated with them) is fully indeterminate.\footnote{A possible alternative would be to consider a dynamical set of events, such that an event that has a totally random propensity associated is not considered as ``existing''.} Worth noticing is the position of certain authors, notably D. Albert \cite{albert}, who have criticized the possibility of basing the distinction between past and future on \emph{records}. This relies on the argument that independently of how much redundancy of information allegedly about the past (i.e. records) one can observe at present, it is always possible to account for it by statistical fluctuations. Albert proposes instead the concept of \emph{intervention}, that is, the belief that the future counterfactually depends on present choices while the past does not (this is so because Albert considers an underlying deterministic description, therefore one can only imagine possible different worlds where a choice could have differed from the actual one). Our model, on the other hand, presupposes the existence of genuinely indeterministic (yet causally related) events and this naturally provides a direction of time. The information about the past --the records in Albert's jargon-- has the same nature of the information about the future, namely both encode the propensities that relate causes and effects. The asymmetry does not arise from different statuses granted to past and future but rather from the amount of information required to encode the causal connection about the past and about the future, respectively.

Finally, the fact that the information recorded in the universe about the causal connection goes farther away in the past than in the future is not the only asymmetry between the past and the future in an indeterministic world. In fact, while future events eventually become determinate (the associated propensities for an event to happen will either become 0 or 1, so every statement will become decidable for a certain time), the information erased from the past is forever lost, and past statements will forever remain undecidable (which here means fundamentally indeterminate). This wass already hinted at by J. Butterfield: ``Dummett’s idea tends to condemn the past to a more endemic unreality [i.e. fundamental indeterminacy (see discussion above)] than the future. For, on the one hand, singular observational statements about the future seem effectively decidable --we naturally envisage making an expedition to the place and time in question, with instruments, if need be, in hand. On the other hand, for analogous statements about the past, there is no such procedure'' \cite{butterfield}. In this sense, the openness of the past becomes wider and wider as time passes.

To conclude, we would like to clarify that the toy model here presented has no claim of being any close to the real mechanism (if any exists) that erases the past as time passes. It is intended as the simplest conceptual way to formalize the idea that the past can be (again) open in an indeterministic world, while maintaining the desired asymmetry between past and future. We hope in this way to stimulate further debate on the topic of indeterminism and its implications on our way of understanding past, present and future in indeterministic theory, possibly within quantum mechanics as well. 

\acknowledgements
We thank Yuval Dolev and Jeremy Butterfield for pointing out relevant literature. F.D.S. acknowledges the financial support  from European Innovation Council and SMEs Executive Agency,
TEQ 766900 / Project OEUP0259 and FWF through the project “Black-box quantum information under spacetime symmetries”, OFWF033730.

\begin{small}

\end{small}


\begin{thebibliography}{200}

\bibitem{norton} Norton, J D. 2003. Causation as Folk Science. \textit{Philosophers' Imprint}. 3 (4): 1–22
\bibitem{gisin1} Gisin, N. (2019). Indeterminism in Physics, classical chaos and Bohmian mechanics. Are real numbers really real?. \emph{Erkenntnis}, 86(6), 1469-1481.
\bibitem{delsantogisin1} Del Santo, F. and Gisin, N. (2019). Physics without determinism: Alternative interpretations of classical physics. \emph{Physical Review A}, 100(6), 062107.
\bibitem{NGHiddenReals} Gisin, N. (2019) {\it Real Numbers as the Hidden Variables of Classical Mechanics}. In \emph{Quantum Studies: Mathematics and Foundations}, 7(2), 197-201.
\bibitem{delsantogisin2} Del Santo, F. and Gisin, N. (2021). The relativity of indeterminacy. \emph{Entropy} 23, 1326.
\bibitem{delsanto2021} Del Santo, F. (2021). Indeterminism, causality and information: Has physics ever been deterministic?. In \emph{Undecidability, Uncomputability, and Unpredictability} (63-79). Springer, Cham.
\bibitem{weizsacker} C. F. von Weizs\"acker. (1971). \emph{Die Einheit der Natur}, München: Hanser.



\bibitem{luka} \L ukasiewicz, J. (1946). On determinism. In McCall, S (ed.). 1967. \emph{Polish Logic, 1920-1939}. Oxford University Press.
\bibitem{dummett68} Dummett, M. (1968). The reality of the past. In \emph{Proceedings of the Aristotelian Society} (Vol. 69, pp. 239-258). Aristotelian Society, Wiley.
\bibitem{dummett04} Dummett, M. (2004). \emph{Truth and the Past}. New York: Columbia University Press.



\bibitem{landauer} Landauer, R. (1996). The physical nature of information. \emph{Physics letters A}, 217(4-5), 188-193.
\bibitem{lombardi} Lombardi, O. (2004). What is information?. \emph{Foundations of Science}, 9(2), 105-134.
\bibitem{floridi} Sequoiah-Grayson, Sebastian and Luciano Floridi. (2005). Semantic Conceptions of Information, \emph{The Stanford Encyclopedia of Philosophy} (Spring 2022 Edition).

\bibitem{popper} Popper, K. R. (1959). The propensity interpretation of probability. \emph{The British Journal for the Philosophy of Science}, 10(37), 25-42.

\bibitem{earman}Earman, J. (1992) Determinism in the Physical Sciences, in Salmon, M. H. (ed.), \emph{Introduction to the Philosophy of Science}. Hackett.

\bibitem{salmon} Salmon, W. C. (1998). Probabilistic Causality, in \emph{Causality and explanation}. Oxford: Oxford University Press.

\bibitem{dowe} Dowe, P. (2000) \emph{Physical Causation: Cambridge Studies in Probability, Induction, and Decision Theory}, Cambridge: Cambridge University Press.


\bibitem{reichenbach} Reichenbach, H. (1956). \emph{The Direction of Time}, Berkeley: University of California Press.

\bibitem{good} Good, I. J. (1961). A Causal Calculus. \emph{British Journal for the Philosophy of Science} 11, 305–318.
\bibitem{suppes2} Suppes, P. (1970). \emph{A Probabilistic Theory of Causality}. Amsterdam: North-Holland.


\bibitem{gisinprop1} Gisin, N. (1984). Propensities and the state‐property structure of classical and quantum systems. \emph{Journal of mathematical physics}, 25(7), 2260-2265.
\bibitem{gisinprop2} Gisin, N. (1991). Propensities in a non-deterministic physics. \emph{Synthese}, 89(2), 287-297.


\bibitem{intuition1} Gisin, N. (2020). Mathematical languages shape our understanding of time in physics. \emph{Nature Physics}, 16(2), 114-116.
\bibitem{intuition2} Gisin, N. (2021). Indeterminism in physics and intuitionistic mathematics. \emph{Synthese}, 1-27.



\bibitem{dummett05} Dummett, M. (2005). The justificationist's response to a realist. \emph{Mind}, 114(455), 671-688.



\bibitem{gisinreal1} Gisin, N. (2012). Non-realism: Deep Thought or a Soft Option?. \emph{Foundations of Physics} 42, 80–85.
\bibitem{gisinreal2} Gisin, N. (2015) A Possible Definition of a Realistic Physics Theory. \emph{Internation Journal of Quantum Foundations}, 1(1), 18-24.

\bibitem{markosian} Markosian, N. (1995). The open past. \emph{Philosophical Studies: An International Journal for Philosophy in the Analytic Tradition}, 79(1), 95-105.

\bibitem{albert} Albert, D. Z. (2000). \emph{Time and chance}. Harvard University Press, Cambrige.

\bibitem{butterfield} Butterfield J. (2012). On time chez Dummett. \emph{European Journal of Analytic Philosophy}, 8(1), 77-102.



\end{thebibliography}
\end{document}